\documentclass[aps,pra,showpacs, superscriptaddress,twocolumn]{revtex4-1}

\usepackage{graphicx,amsmath,hyperref}
\usepackage{color}
\usepackage{soul}
\usepackage{braket}
\usepackage{bm} % Bolding of greek letters

\begin{document}

%\preprint{}

\title{Complete positivity, finite temperature effects, and additivity of noise for time-local qubit dynamics}

\author{Juho Lankinen} 
\affiliation{Turku Center for Quantum Physics, Department of Physics and
Astronomy, University of Turku, FIN-20014 Turku, Finland}

\author{Henri Lyyra} 
\affiliation{Turku Center for Quantum Physics, Department of Physics and
Astronomy, University of Turku, FIN-20014 Turku, Finland}

\author{Boris Sokolov} 
\affiliation{Turku Center for Quantum Physics, Department of Physics and
Astronomy, University of Turku, FIN-20014 Turku, Finland}

\author{Jose Teittinen} 
\affiliation{Turku Center for Quantum Physics, Department of Physics and
Astronomy, University of Turku, FIN-20014 Turku, Finland}

\author{Babak Ziaei} 
\affiliation{Turku Center for Quantum Physics, Department of Physics and
Astronomy, University of Turku, FIN-20014 Turku, Finland}

\author{Sabrina Maniscalco} 
\affiliation{Turku Center for Quantum Physics, Department of Physics and
Astronomy, University of Turku, FIN-20014 Turku, Finland}
\email[]{smanis@utu.fi} \homepage[]{www.openquantum.co.uk}

\date{\today}

\begin{abstract}
We present a general model of qubit dynamics which entails pure dephasing and dissipative time-local master equations. This allows us to describe the combined effect of thermalisation and dephasing beyond the usual Markovian approximation. We investigate the complete positivity conditions and introduce a heuristic model that is always physical and provides the correct Markovian limit. We study the effects of temperature on the non-Markovian behaviour of the system and show that the noise additivity property discussed by Yu and Eberly in Ref. \cite{YuEberly} holds beyond the Markovian limit.

\end{abstract}

%\pacs{03.65.Ta, 03.65.Yz, 03.75.Gg}

\maketitle

\section{Introduction}

Noise induced by the environment has been for decades considered the major enemy of quantum technologies. It is nowadays recognised that this initial belief was wrong \cite{Verstraete}. Not only  can noise be used to generate quantum properties such as entanglement \cite{Braun,Zell,Barreiro,Mazzola}, but also the dynamics of an open system, e.g., its coherence time, can be modified by reservoir engineering. Generally, there are a number of ways to change the properties of the environment in a selective and controllable way. Typical examples are modifications of the spectral density of the electromagnetic field acting as an environment, such as in cavity quantum electrodynamics or photonics band gap materials \cite{john,lambropoulos}, and dynamical decoupling methods \cite{ref1Lorenza, ref2Lorenza,NMDD1,NMDD2}. These techniques are nowadays routinely performed in the laboratories \cite{DDengineer}. 

At the same time, the experimental ability to isolate from the environment and coherently control individual qubits in solid state systems such as NV centres in diamonds \cite{NVCentres} and superconducting Josephson junctions \cite{Squid} has made them ideal candidates for quantum technologies. However, despite the advances of the last decade, the effects of noise in these systems still needs to be taken into account to study their robustness, efficiency, and life-time in realistic physical conditions.

The rising importance of both reservoir engineering techniques and solid state qubits highlights the need to investigate open quantum systems models which go beyond the  Markovian approximation usually used in  quantum optics. During the last few years research on non-Markovian dynamics has flourished. The study of memory effects, characterising non-Markovian systems, has been linked with a partial revival of information on the open system \cite{BLP,NMNatPhys}. Several ways to quantify information flow and back-flow have been proposed in order to understand the physical phenomena underlying non-Markovian evolution \cite{NMQJ,NMWolf,NMRHP,NMCV,NMFisher,NMLuo,NMFrancesco,NMSabDarek}. Finally, intense research activity is currently focussed on the understanding of the conditions and the potential advantages of memory effects to enhance the performance of quantum devices \cite{NMQCAP,NMRQKD,NMRMetrology,NMRTeleportation, Andrea}.

The main difficulty when dealing with non-Markovian models is the lack of a general theorem which guarantees the physicality of the state as time evolves. From a mathematical point of view, one of the reasons why Markovian master equations have been so popular is indeed the existence of the Gorini-Kossakowski-Sudarshan-Lindblad  (GKSL) theorem characterising completely positive and trace preserving (CPTP) dynamical maps \cite{Lindblad,GKS}. This in turn guarantees that, in absence of initial system-environment correlations,  the time evolution of any quantum state of the open system, as described by the solution of the GKSL master equation, is always physical. 

Because of this difficulty, dealing with generalised non-Markovian master equations is always a tricky business \cite{Hazards,Laura,Sab1,Sab2}. Even for a single qubit, where conditions for complete positivity are known \cite{Ruskai,algoet}, all studies of non-Markovianity have mostly focused on very simple models for which an exact solution of the total, i.e., system plus environment, dynamics is available \cite{ReviewBLP,ReviewRHP}. This indeed guarantees physicality by construction. Typical examples are the purely dephasing model \cite{puremodel1,puremodel2,puremodel3,Greg,Tomi,PinjaJohn}, the Pauli channel model \cite{Chruscinski2013} and the amplitude damping model \cite{Barry,MazMan}. The latter one goes beyond unital dynamics but is restricted to the case in which the two-state system dissipatively interacts with a zero-temperature reservoir.

In this paper we go beyond the existing literature in several ways. Firstly, we solve and study the CPTP conditions of a generic time-local master equation which contains heating, dissipation and pure dephasing terms. This allows to assess the question of additivity of noise under non-Markovian dynamics, extending the results of Ref. \cite{YuEberly}. Secondly, we discuss the effects of temperature in a non-unital model which, in the Markovian limit, gives the standard Markovian master equation for a two-level atom interacting with a thermal bath. Finally, we show that, as one might expect, the occurrence of non-Markovianity now has a more complicated origin being linked to both the dephasing and the dissipative terms. 

The paper is structured as follows. In Sec. II we introduce the general time-local master equation, present its solution, and show that the noise additivity property holds beyond the Markovian approximation. In Sec. III we study the complete positivity conditions, while in Sec. IV we introduce a heuristic master equation which is always physical and discuss the interplay between temperature effects and non-Markovianity. Finally in Sec. V we present conclusions.

%
%	In section II, we introduce the master equation for our model which is a two-level system and calculate explicit dynamics in the Bloch vector representation for an arbitrary initial state. After that we write the dynamics of the Bloch vector using damping matrix and translation vector to obtain the parameters required for the CP conditions. In section III, we study the complete positivity conditions for three different cases: The general form of time dependent coefficients, short time behavior of the system with and without weak coupling approximation. 

\section{The model}
\subsection{The master equation}
Let us consider the following time-local master equation for the qubit density matrix $\rho$ in the interaction picture and in units of $\hbar$,
\begin{eqnarray}
\frac{d\rho}{dt} &=&L_t (\rho) \equiv - i \omega(t)\left[ \sigma_z, \rho \right] + \frac{\gamma_1 (t)}{2} L_1 (\rho) + \frac{\gamma_2 (t)}{2} L_2(\rho) \nonumber \\
&+& \frac{\gamma_3(t)}{2} L_3(\rho), \label{eq:ME}
\end{eqnarray}
where $\gamma_i(t)$ are time-dependent rates, $\omega(t)$ is a time-dependent frequency shift, and the dissipators $L_i(\rho)$ are defined as
\begin{align}
L_1(\rho) &= \sigma_+ \rho \sigma_- - \frac{1}{2} \left\{ \sigma_-\sigma_+, \rho \right\}, \\
L_2(\rho) &= \sigma_- \rho \sigma_+ - \frac{1}{2} \left\{ \sigma_+\sigma_-, \rho \right\}, \\
L_3(\rho) &= \sigma_z \rho \sigma_z - \rho.
\end{align}
In the equations above $\sigma_{\pm}=\sigma_x \pm i \sigma_y$ are the inversion operators and $\sigma_i$ ($i={x,y,z}$) are the Pauli operators. 
The three dissipators $L_1$, $L_2$, $L_3$, describe heating, dissipation and dephasing, respectively. However, contrarily to the typical GKSL master equation \cite{Lindblad,GKS}, the decay rates are not positive constants but time-dependent functions which need not be positive at all times. The master equation \eqref{eq:ME} describes phase covariant noise and has been considered recently in the context of quantum metrology in noisy channels \cite{Andrea}.

Special cases of master equations of the form of Eq. (\ref{eq:ME}) are those considered, e.g., in Refs. \cite{BLP,NMQJ,NMQCAP,NMRMetrology},  for $\gamma_1(t)=\gamma_3(t)=0$ describing an amplitude damping model, and the pure dephasing master equation considered, e.g., in Refs. \cite{NMNatPhys,NMQCAP,NMRMetrology,Greg,Tomi,PinjaJohn}  for $\gamma_1(t)=\gamma_2(t)=0$. These two special cases can be derived by means of an exact approach starting from a microscopic Hamiltonian model for system and environment. Hence the resulting dynamics is always CPTP. In the more general case considered in this paper, however, the master equation is introduced phenomenologically, since an exact microscopic derivation is unfeasible. As a consequence, restrictions on the form of the time-dependent decay rates arise in order to preserve the CPTP character of the dynamics.

The master equation \eqref{eq:ME} is one of the most general time-local master equations for a qubit. Indeed, it combines the effects of pure dephasing terms and dissipative terms. The dynamics is non-unital and the heating term $L_1$ accounts for the presence of a finite temperature environment. The corresponding dynamical map can be written as $\Phi_t = T \exp\left( -\int_0^t L_s ds \right)$, with $T$ the chronological ordering operator. Whenever one of the time-dependent rates takes negative values then the dynamical map is not CP-divisible, i.e., the propagator $\Lambda_{t,s}$ defined by $\Phi_t=\Lambda_{t,s} \Phi_s$, with $s \le t$, is not CP. In the following we define as Markovian a dynamics such that $\Lambda_{t,s}$ is CP $\forall t,s$.

% Further, the divisibility of the master equation is satisfied if and only if all of the time-dependent coefficients $\gamma(t)$ are non-negative, i.e. $\gamma_i(t)\geq 0, \forall i $.

\subsection{The solution}
Let us indicate with $\ket{1}$ and $\ket{2}$ the ground and excited states of the qubit, respectively. From Eq. \eqref{eq:ME} one straightforwardly derives the following equations for the ground state probability $P_1(t) = \braket{1|\rho(t)|1} $ and the coherence $\alpha (t)= \braket{1|\rho(t)|2}$:
\begin{eqnarray}
\frac{dP_1}{dt}&+&\frac{\gamma_1(t)+\gamma_2(t)}{2} P_1(t)=\frac{\gamma_2(t)}{2}, \\
\frac{d \alpha}{dt}&=& \alpha(t)\left[ 2i\omega(t) + \frac{1}{2} \left(\frac{\gamma_1(t) + \gamma_2(t)}{2} + 2 \gamma_3(t)\right)\right].
\end{eqnarray}
The equations above are linear first-order differential equations and can be solved for any values of the time-dependent decay rates. The solution reads as follows:
\begin{eqnarray}
P_1(t) &=&   e^{-\Gamma(t)}  \left[ G(t) + P_1(0)\right], \label{eq:P1}\\
\alpha (t)&=&   \alpha(0) e^{i\Omega(t) - \Gamma(t)/2 - \tilde{\Gamma}(t)}, \label{eq:alfa}
 \end{eqnarray}
where
\begin{eqnarray}
\Gamma(t) &=& \int_0^t  dt'  [\gamma_1(t') + \gamma_2(t') ]/2, \label{eq:Gamma} \\
\tilde{\Gamma}(t) &=& \int_0^t dt' \gamma_3(t'),  \label{eq:Gammat}\\
\Omega(t)&=& \int_0^t dt' \omega(t'), \label{eq:Omega}  \\
G(t)&=& \int_0^t  dt' e^{\Gamma(t')}  \gamma_2(t') /2. \label{eq:G} 
 \end{eqnarray}
 
 If the time-dependent coefficients quickly attain a stationary positive constant value, after an initial short time interval $\tau_c$, known as the correlation time of the environment, one obtains the approximated GKSL master equation by coarse-graining over $\tau_c$ and extending to infinity the limit of integration in Eqs.  \eqref{eq:Gamma} - \eqref{eq:Omega}. More precisely one obtains the following Markovian limits for the quantities defined in Eqs. \eqref{eq:Gamma} - \eqref{eq:G}.
\begin{eqnarray}
\Gamma_M &=& (\gamma_1 + \gamma_2 )t /2, \\
\tilde{\Gamma}_M &=& \gamma_3 t, \\
\Omega_M&=& \omega t,  \\
G_M&=& \frac{\gamma_2}{\gamma_1+\gamma_2} \left(e^{(\gamma_1+\gamma_2)t/2} -1 \right).
 \end{eqnarray}  
Using these expressions one can recover the well known Markovian formulas for populations and coherences
\begin{eqnarray}
P_1(t) &=&   e^{-(\gamma_1+\gamma_2)t/2} P_1(0)   +  \frac{\gamma_2}{\gamma_1+\gamma_2} \left( 1- e^{-(\gamma_1+\gamma_2)t/2}\right), \\
\alpha (t)&=&   \alpha(0) e^{i\omega t- (\gamma_1 + \gamma_2 )t /4 - \gamma_3 t}. \label{eq:cohsol}
 \end{eqnarray}
The approximated GKSL master equation, obtained from Eq. \eqref{eq:ME} by simply replacing the time-dependent coefficients with the corresponding positive constants, has been investigated, e.g., in Ref. \cite{YuEberly} to study additivity of noise in the Markovian limit and for $T=0$. There the authors show that, while for a single qubit additivity holds, composite systems may violate this property. In the single qubit case, additivity simply means that the decay rates of the off-diagonal elements of the density matrix, when the qubit is subjected to independent sources of noise, is just the sum of the decay rates arising from the interaction with each individual environment. This is straightforwardly seen in Eq. \eqref{eq:cohsol}. 

In this paper we generalise the results presented in Ref. \cite{YuEberly} to the case of general temperatures and beyond the Markovian approximation. Equation \eqref{eq:alfa}, indeed, straightforwardly proves that additivity holds for the general time-local master equation \eqref{eq:ME}, provided that the solution is physical. In the following section we will thoroughly investigate the conditions under which the master equation \eqref{eq:ME} gives rise to a physically admittable dynamics described by a CPTP map and we will give examples of both physical and unphysical behaviour for specific choices of the time-dependent decay rates.

\section{Complete positivity}
Let us begin by expressing the solution in terms of components of the Bloch vector defined by $\rho(t) = \frac{1}{2} \left(I + \mathbf{ v} \cdot \bm{\sigma}\right)$, with $I$ the identity operator, $\bm{\sigma}$ the Pauli operators vector having as components $\sigma_i$, ($i=x,y,z$), and $\mathbf{v}=(x_1,x_2,x_3)$ the Bloch vector. The evolution of the latter one is given by
\begin{eqnarray}
\mathbf{ v}(t) = \Lambda (t) \mathbf{ v}(0) + \mathbf{ T}(t),
 \end{eqnarray}
where $\Lambda$ is known as the damping matrix and $\mathbf{T}(t)=(0,0,t_3(t))$ is the translation vector given by
\begin{eqnarray}
t_3(t) =  e^{-\Gamma(t)} [1+2 G(t)] - 1. \label{eq:t3}
 \end{eqnarray}
The eigenvalues of the damping matrix can be written as
\begin{eqnarray}
\lambda_1(t) &= &e^{-\Gamma(t)/2 - \tilde{\Gamma}(t) +i \Omega(t)}, \\
 \lambda_2(t) &= &e^{-\Gamma(t)/2 - \tilde{\Gamma}(t)- i \Omega(t)}, \\
 \lambda_3(t) &= &e^{-\Gamma(t)}. \label{eq:lambda3}
 \end{eqnarray}

\subsection{CP criteria}

Complete positivity conditions can be expressed in terms of inequalities involving the Bloch vector components \cite{Ruskai}. In the following we will use the formulation introduced in Ref. \cite{algoet} 
\begin{eqnarray}
|p(t)|, |q(t)| &\le& \frac{1}{2}, \label{eq:cp1} \\
y(t)^2 &\le& \left( \frac{1}{2} - p(t) \right)\left( \frac{1}{2} + q(t) \right), \\
w(t)^2 &\le &\left( \frac{1}{2} - q(t) \right)\left( \frac{1}{2} + p(t) \right), \label{eq:cp3}
 \end{eqnarray}
where
\begin{eqnarray}
p(t)=\frac{1}{2}[t_3(t)+\lambda_3(t)], \\
q(t)=\frac{1}{2}[t_3(t)-\lambda_3(t)], \\
w(t)=\frac{1}{2}[\lambda_1(t)+\lambda_2(t)], \\
y(t)=\frac{1}{2}[\lambda_1(t)-\lambda_2(t)].
 \end{eqnarray}
 
%For our system these four equations read as
%\begin{eqnarray}
%p(t) &=& e^{-\Gamma (t)} [G(t)+\frac{1}{2}] - \frac{1}{2} + \frac{e^{-\Gamma (t)}}{2} ,  \label{eq:p} \\ 
%q(t) &=& e^{-\Gamma (t)} [G(t)+\frac{1}{2}] - \frac{1}{2} - \frac{e^{-\Gamma (t)}}{2}, \\
%y^2(t) &=& - \sin^2 \Omega(t) e^{-\Gamma(t) - 2 \tilde{\Gamma}(t)}, \\
%w^2(t) &=&  \cos^2 \Omega(t) e^{-\Gamma(t) - 2 \tilde{\Gamma}(t)}. \label{eq:w}
% \end{eqnarray}

Using the analytical expressions given by Eqs. \eqref{eq:t3}-\eqref{eq:lambda3}, the CP necessary and sufficient conditions read as follows
\begin{eqnarray}
i) && 0 \leq e^{-\Gamma (t)} \left( G(t)+1 \right) \leq 1, \label{pcond}\\
ii) && 0 \leq e^{-\Gamma (t)} G(t) \leq 1, \label{qcond} \\
iii)  &&- e^{-\Gamma(t)-2\tilde{\Gamma}(t)}\!\sin^2 \Omega(t)\! \leq \!e^{-\Gamma(t)}G(t) \!\! \left[ \!1 - e^{-\Gamma(t)} (G(t)+1) \!\right]\!\!, \nonumber \\ \label{ycond} \\
iv)  && e^{-\Gamma(t)-2\tilde{\Gamma}(t)} \! \cos^2 \Omega(t)\! \leq \! e^{-\Gamma(t)} \! \left[ \!1 - e^{-\Gamma(t)}G(t) \right] [ G(t) + 1]. \nonumber \\\label{wcond} 
\end{eqnarray}
We notice that the validity of conditions $i)-ii)$ (positivity conditions) implies that condition $iii)$ is always satisfied, as the l.h.s. of the inequality is always non-positive and the r.h.s. is always non-negative. We also stress that the dephasing term described by $L_3$ directly influences only conditions $iii)-iv)$ via the decoherence term $\tilde{\Gamma}(t)$. Finally we note that $\Gamma(t)\ge 0$ and $1\ge G(t)\ge 0$ are sufficient conditions for positivity, i.e., for $i)-ii)$. 

Let us focus on the case in which the purely dephasing term vanishes, namely $\gamma_3(t)=0$. In this case one sees that condition $iv)$ simplifies and can be recast as follows
\begin{equation}
e^{-\Gamma(t)}  \cos^2\Omega(t) \le e^{-\Gamma(t)} + e^{-\Gamma(t)} G(t) [1-e^{-\Gamma(t)} (G(t)+1)]. \label{eq:ineq4}
\end{equation}
For $\Omega(t)=0$ one sees immediately that the inequality above is automatically satisfied whenever the positivity conditions $i)-ii)$ are satisfied. In the more general case in which $\Omega(t)\neq 0$, the condition is still valid provided the positivity conditions hold since, at any time, the left hand side of the inequality \eqref{eq:ineq4} is upper bounded by $e^{-\Gamma(t)}$.

From the reasoning above one can reach a simple conclusion regarding the physicality of the general form of master equation \eqref{eq:ME}. Indeed in this case, assuming that conditions $i)-ii)$ are verified,  a sufficient condition for complete positivity is that the term $\tilde{\Gamma}(t) \ge 0$.

\subsection{Weak coupling and short-time limits}
We conclude this section by looking at the weak coupling and short time limits.
In the weak coupling limit the following approximations hold
\begin{eqnarray}
e^{-\Gamma (t)} \left( G(t)+1 \right) &\simeq& 1-\int_0^t \gamma_1(t') dt', \\
e^{-\Gamma (t)} G(t) &\simeq& \int_0^t \gamma_2(t') dt'.  \\
\end{eqnarray}
Using these approximations a straightforward calculation shows that conditions $i)-ii)$ and $iv)$ correspond to
\begin{eqnarray}
i) \; && \int_0^t \gamma_2(t') dt' \ge 0, \\
ii) \; &&  \int_0^t \gamma_1(t') dt' \ge 0, \\
iv) \; && \int_0^t \gamma_3(t') dt' \ge 0.
\end{eqnarray}

As for the short-time approximation, by considering the Taylor expansion around $t=0$ of the exponential terms, i.e., $e^{-\Gamma(t)}$ and of the term $e^{-\Gamma(t)} G(t)$, it is easy to convince oneself that the CP conditions amount at $i)  \; \gamma_1(0) \ge 0$, $ii)  \; \gamma_2(0) \ge 0$, $iv) \; \gamma_3(0) \ge 0$. These conditions imply that, contrarily to what happens in certain unital time-local master equations (see, e.g., Ref. \cite{Hall2014}), in our model the decay rates cannot take at the initial time negative values.

\section{Thermal effects and non-Markovianity}

Let us now consider the following heuristic model. We assume that the open quantum system of interest is coupled to both a thermal reservoir and a dephasing environment at the same temperature $T$. The former one induces heating and dissipation at rates given by $\gamma_1(t)/2= N f(t)$ and $\gamma_2(t)/2=(N+1) f(t)$, with $N$ the mean number of excitations in the modes of the thermal environment. We notice that, for a zero $T$ environment, the heating rate $\gamma_1(t)=0$ while the dissipation rate $\gamma_2(t)=f(t)$. Hence, we consider as a possible physically reasonable choice for the time-dependent function $f(t)$ the one obtained in the exactly solvable zero-$T$ model presented, e.g., in Ref. \cite{MazMan}. In this model the function $f(t)$ takes the form
\begin{equation}
f(t)=-2 {\rm Re} \left\{ \frac{\dot{c}(t)}{c(t)}\right\}, \label{eq:ft}
\end{equation}
with
\begin{equation}
c(\tau)= e^{-\tau/2} \left[ \cosh (d\tau/2)  + \frac{\sinh (d\tau/2)}{d}  \right] c(0), 
\end{equation}
where $d=\sqrt{1-2R}$, and $R$ is a dimensionless positive number measuring the overall coupling between the two-state system and the environment with respect to the width of the spectral density of the environment.
The coefficient $\Gamma(t)$ can be analytically calculated and yields the simple expression
\begin{eqnarray}
\Gamma(t)&=&(2N+1) \int_0^t f(t') dt' =- \ln \left[ \left(\frac{c(t)}{c(0)}\right)^{2(2N+1)}\right] \nonumber \\ 
&\equiv& -\ln [x(t)^{2N+1}], \label{eq:example}
\end{eqnarray}
where we have used Eq. \eqref{eq:ft} and defined $x(t)=[c(t)/c(0)]^2$. We note that $0 \le x(t) \le 1$ and that $x(t)$ presents oscillations in time only for $R > 1/2$ (strong coupling, broad spectral density) while it decays monotonically for $R< 1/2$ (weak coupling,  narrow spectral density).  It is straightforward to see by explicitly calculating the decay rates $\gamma_1(t)$ and $\gamma_2(t)$ that they are always positive whenever $R<1/2$ (divisible dynamics) and attain temporarily negative values for $R>1/2$ (non divisible dynamics). 

\begin{figure}[!t]
\centering
\includegraphics[scale=0.5]{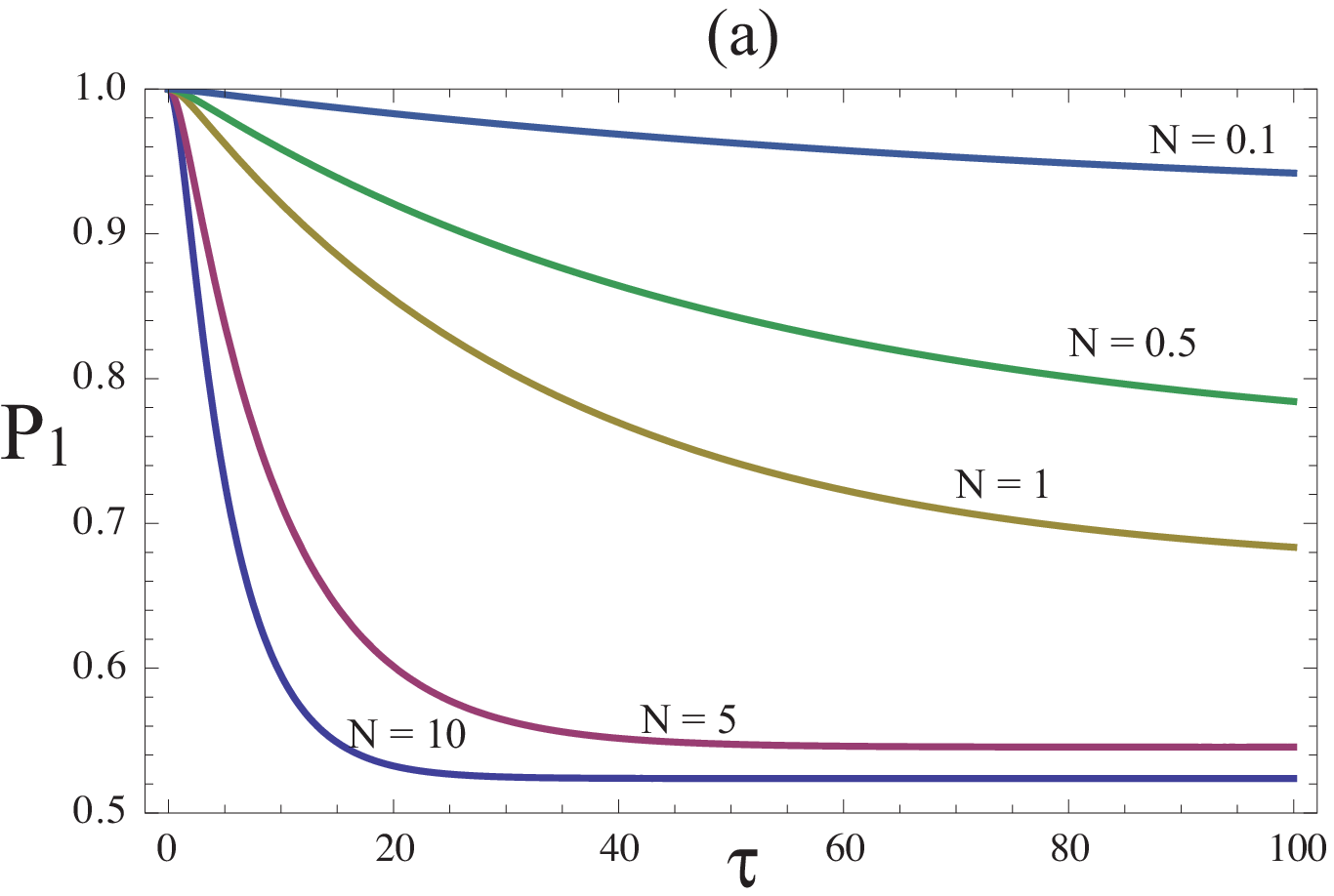}
\hspace{1cm}
\includegraphics[scale=0.5]{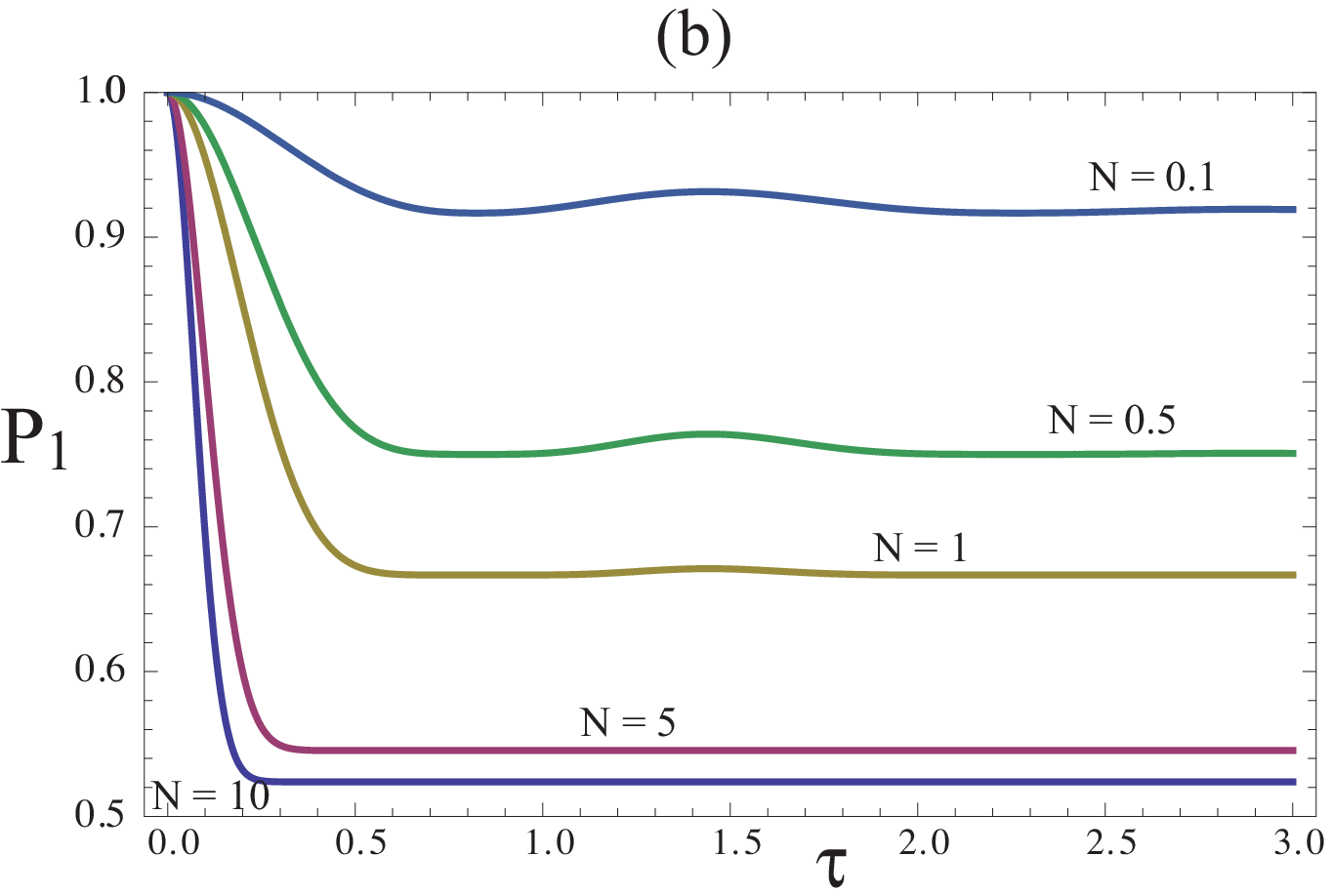}
\caption{(Color online) Dynamics of the ground state population (a) in the weak coupling (Markovian) regime, $R=0.01$, and (b) in the strong coupling (non-Markovian) regime, $R=10$, for different values of $N$ (temperature).}
\label{A1}
\end{figure}

By inserting Eq. \eqref{eq:example} into Eq. \eqref{eq:P1} one obtaines the following analytic expression for the ground state population 
\begin{equation}
P_1(t) = x(t)^{2N+1} P_1(0) + \frac{N+1}{2N+1} \left(1- x(t)^{2N+1} \right)
\end{equation}
It is straightforward to verify that, for this model, the positivity conditions $i)-ii)$ are verified at all times and for all values of $R>0$. We notice that, in absence of the pure dephasing term, i.e., whenever $\gamma_3(t)=0$, condition $iv)$ is automatically satisfied and the corresponding $T$-temperature master equation is always physical. Moreover, not only this model by construction reduces to the exact zero-$T$ model, but it also gives the correct Markovian limit for a two-level system in a thermal bath at $T$ temperature. Indeed, if we indicate with $\gamma_M$ the Markovian limit of $\gamma_2(t)$ in the exact zero-$T$ model, one can easily see that, the Markovian expressions of the decoherence factor $e^{-\Gamma(t)}$ and of the ground state probability $P_1(t)$, obtained for $R\ll1$, read  
\begin{eqnarray}
e^{-\Gamma(t)} &\rightarrow& e^{-(N+1/2) \gamma_M t} \nonumber \\
P_1(t) &\rightarrow& e^{-(N+1/2) \gamma_M t} \left[P_1(0) + \frac{N+1}{2N+1}(1-e^{-(N+1/2) \gamma_M t}) \right], \nonumber
\end{eqnarray}
respectively.

We now go back to the situation in which the pure dephasing term is present in the Liouvillian. The coefficient $\gamma_3(t)$ does not influence the behaviour of the populations.
In Fig. 1 we plot the time evolution of the ground state population $P_1(t)$ as a function of time for different temperatures, i.e. $N$, in both the Markovian case, Fig. 1 (a), and the non-Markovian case, Fig. 1 (b). We notice that, for $R\gg1$ and for increasing values of temperature, the oscillations in ground state population are quickly damped, even if the dynamics continues to be non-Markovian because both the $\gamma_1(t)$ and the $\gamma_2(t)$ decay rates take negative values. Hence, the presence of oscillations in the ground or excited state probability is not just connected to the Markovian or non-Markovian character of the dynamics, as it was for the exact model of Ref. \cite{MazMan}, but depends also on the temperature of the environment. 

We now consider the effect of the pure dephasing term. As done before, we will again use a model of pure dephasing which arises from an exact microscopic description \cite{puremodel1,puremodel2,puremodel3}. In this case the analytic expression for the dephasing rate is given by
\begin{equation}
\gamma_3(t) = 2\int d \omega J(\omega) \coth (\omega / k_B T) \sin(\omega_c t),
\end{equation}
where $k_B$ is the Boltzmann constant and the spectral density is assumed to be of the Ohmic class
\begin{eqnarray}
J(\omega)= \alpha \frac{\omega^s}{\omega_c^s} e^{-\omega/\omega_c} ,
\end{eqnarray}
with $\alpha$ an overall coupling constant and $\omega_c$ the cutoff frequency.

It is worth stressing that this model always leads to $\tilde{\Gamma} (t) \ge 0$, hence the dynamics is not only positive but also completely positive since condition $iv)$ is verified at all times. In Ref. \cite{Tomi} the non-Markovianity of this model was studied in detail and was found to be linked to the value of the Ohmicity parameter $s$. Hence, the two parameters governing the Markovian to non-Markovian crossover are $R$ and $s$. In other words, the dynamics of the whole system can be non-Markovian also for values of $R \le 1/2$ provided that the Ohmicity parameter is such that $\gamma_3(t) <0$ for certain time intervals.

Generally, the dynamics of the coherences, given by Eq. \eqref{eq:cohsol}, will be damped because of both the heating/dissipation terms and the dephasing term. We notice however, that some of the characteristic phenomena typical of pure dephasing in Ohmic-like environments, e.g., coherence trapping \cite{Tomi}, will not occur in this model because the coefficient $e^{- \Gamma(t)}$ will always eventually erase the coherences and drive the system towards a thermal mixed state.

\section{Conclusions}

We have solved and investigated the dynamics of a general time-local master equation which combines dissipative and pure dephasing terms showing that decoherence is still additive. Guided from the knowledge of the exact microscopic  amplitude damping and pure dephasing master equations, we have introduced an intuitive heuristic model which is always CPTP. This model allows us to study the effects of finite temperature on the dynamics of the qubit in the non-Markovian regime. As expected, when increasing the temperature of the environment, decay of both populations and coherences is faster and faster. Moreover, thermalisation destroys phenomena such as coherence trapping which are present in purely dephasing systems. We have pointed out that the conditions for non-Markovian dynamics are now dependent on both the characteristic parameter of the dissipative terms, $R$, and on the corresponding parameter for the pure dephasing term, $s$. Finally we have seen that finite temperature effects quickly destroy the oscillatory behaviour of populations even in the strongly non-Markovian regime $R\gg1$.

Given the importance of studies of fundamental non-Markovian models, we believe that our results will be of use for both reservoir engineering and to model noise in solid-state devices in realistic experimental conditions, i.e., when finite-temperature effects cannot be neglected. As an example, an interesting future direction is the investigation of whether and how memory effects may affect the break down of additivity property in bipartite systems, as it happens in the Markovian case \cite{YuEberly}.

\section*{Acknowledgements} 
This work was supported by  the EU Collaborative project QuProCS (Grant Agreement 641277), the Academy of Finland (Project no. 287750) and the Magnus Ehrnrooth Foundation.


\begin{thebibliography}{30}

\bibitem{YuEberly} 
T. Yu and J. H. Eberly, Phys. Rev. Lett. {\bf 97}, 140403 (2006).

\bibitem{Verstraete} F. Verstraete, M. M. Wolf and J. I. Cirac, Nature Physics {\bf 5}, 633 (2009).

\bibitem{Braun} D. Braun, Phys. Rev. Lett. {\bf 89}, 277901, (2005).
\bibitem{Zell} T. Zell, F. Queisser, and R. Klesse, Phys. Rev. Lett. {\bf 102}, 160501 (2009).
\bibitem{Barreiro} J. T. Barreiro {\it et al.}, Nat. Phys. {\bf 6}, 943 (2010).
\bibitem{Mazzola} L. Mazzola, S. Maniscalco, J. Piilo, and K.-A. Suominen, Phys. Rev. A {\bf 79} 042302 (2009).


\bibitem{john}
S. John, and T. Quang, Phys. Rev. A
{\bf 50,} 1764 (1994).

\bibitem{lambropoulos}
P. Lambropoulos {\it et al.}, Rep. Prog. Phys. {\bf 63}, 455 (2000).

\bibitem{ref1Lorenza} L. Viola and S. Lloyd, Phys. Rev. A {\bf 58} 2733 (1998 ).%1
\bibitem{ref2Lorenza} L. Viola, E. Knill  and S. Lloyd, Phys. Rev. Lett. {\bf 82} 2417 (1998). %2

\bibitem{NMDD1} A. G. Kofman and G. Kurizki G,  Phys. Rev. Lett. {\bf 87} 270405 (2001).
\bibitem{NMDD2} A. G. Kofman and G. Kurizki, Phys. Rev. Lett. {\bf 93} 130406 (2004). 

\bibitem{DDengineer} 
M. J. Biercuk,  {\it et. al.},  Nature {\bf 458}, 996-1000 (2009).

\bibitem{NVCentres}
M. W. Doherty {\it et al.}, Physics Reports {\bf 528}, 1 (2013).

\bibitem{Squid}
J. Q. You and F. Nori, Nature {\bf 474}, 589 (2011).

\bibitem{BLP} H.-P. Breuer, E.-M. Laine, and J. Piilo, Phys. Rev. Lett. {\bf 103}, 210401 (2009).

\bibitem{NMNatPhys}
B-H. Liu,  {\it et al}.
 Nature Phys.
{\bf 7}, 931-934 (2011).

\bibitem{NMQJ}
J. Piilo, S. Maniscalco, K. H\"{a}rk\"{o}nen, and K.-A.Suominen, 
 Phys. Rev. Lett. {\bf 100}, 180402 (2008).

\bibitem{NMWolf}
M. M. Wolf, J. Eisert, T. S. Cubitt,  and J. I. Cirac, 
Phys. Rev. Lett.
{\bf 101}, 150402 (2008).


\bibitem{NMRHP}
A. Rivas, S. F. Huelga,  and M. B. Plenio, 
 Phys. Rev. Lett. {\bf 105}, 050403 (2010).

\bibitem{NMCV}
R. Vasile, S. Maniscalco, M. G. A.Paris, H-P. Breuer, and J. Piilo J, Phys. Rev. A
{\bf 84}, 052118 (2011).

\bibitem{NMFisher} X.-M. Lu,  X., Wang,  and  C.P. Sun, Phys. Rev. A {\bf 82}, 042103 (2010).

\bibitem{NMLuo}
S. Luo, S. Fu, and H. Song, Phys. Rev. A
{\bf 86}, 044101 (2012).


\bibitem{NMFrancesco}
S. Lorenzo, F. Plastina, and M.  Paternostro, 
 Phys. Rev. A
{\bf 88}, 020102(R) (2013).

\bibitem{NMSabDarek}
D. Chru\'{s}ci\'{n}ski, and S. Maniscalco, 
Phys. Rev. Lett.
{\bf 112}, 120404 (2014).

\bibitem{NMQCAP}
B. Bylicka, D. Chru\'sci\'nski, and S. Maniscalco,  Scientific Reports
{\bf 4} 5720 (2014).

\bibitem{NMRQKD}
R. Vasile, S. Olivares, M. G. A.  Paris, and S. Maniscalco, Phys. Rev. A
{\bf 83}, 042321 (2011).

\bibitem{NMRMetrology}
A. W. Chin, S. F. Huelga, and M. B. Plenio, Phys. Rev. Lett.
{\bf 109}, 233601 (2012).

\bibitem{NMRTeleportation}
E-M. Laine, H-P. Breuer, and J. Piilo 
{\it Scientific Reports}
{\bf 4}, 4620 (2014).

\bibitem{Andrea} A. Smirne, J. Kolodynski, S. F. Huelga, and R. Demkowicz-Dobrzanski, arXiv:1511.02708.

\bibitem{Lindblad} G. Lindblad,  Commun. Math. Phys. {\bf 48}, 119 (1976). 
\bibitem{GKS}  V. Gorini,  A. Kossakowski,and E.C.  Sudarshan,  J. Math. Phys. \textbf{17}, 821 (1976). 



\bibitem{Hazards} S. M. Barnett and S. Stenholm, Phys. Rev. A {\bf 64}, 033808
(2001).

\bibitem{Laura} L. Mazzola, E.-M. Laine, H.-P. Breuer, S. Maniscalco, and J. Piilo, Phys. Rev. A {\bf 81}, 062120 (2010).

\bibitem{Sab1} S. Maniscalco Phys. Rev. A {\bf 75}, 062103 (2007).

\bibitem{Sab2} S. Maniscalco and F. Petruccione, Phys. Rev. A {\bf 73}, 012111 (2006).

\bibitem{Ruskai} M. B. Ruskai, S. Szarek, and E. Werner, Linear Algebr. Appl. {\bf 347}, 159 (2002).

\bibitem{algoet} A. Fujiwara and P. Algoet, Phys. Rev. A {\bf 59}, 3290 (1999).

\bibitem{ReviewBLP}
H-P. Breuer, E-M. Laine, and J. Piilo, and B. Vacchini, 
arXiv:1505.01385
(2015).

\bibitem{ReviewRHP}
 A. Rivas, S. F. Huelga,  and M. B.  Plenio, Rep. Prog. Phys. {\bf  77}, 094001 (2014).

\bibitem{puremodel1} J. Luczka  Physica A {\bf 167}, 919 (1990). %Exactly Solvable
\bibitem{puremodel2} G. M. Palma, K.-A. Suominen, and A.-K. Ekert , Proc. Roy. Soc. Lond. A {\bf 452}, 567 (1996). %Dephasing Model 
\bibitem{puremodel3} J. Q. Reinga, L. Quiroga, and N. F. Johnson  Phys. Rev. A. {\bf 65} 032326 (2002).


\bibitem{Tomi} P. Haikka, T. H. Johnson, and S. Maniscalco
Phys. Rev. A {\bf 87}, 010103(R) (2013).

\bibitem{Greg} C. Addis, G. Brebner, P. Haikka and S. Maniscalco Phys. Rev. A {\bf 89}, 024101 (2014).

\bibitem{PinjaJohn}
P. Haikka, J. Goold, S. McEndoo, F. Plastina, and S. Maniscalco, Phys. Rev. A
{\bf 85}, 060101(R) (2012).

\bibitem{Chruscinski2013}
D. Chru\'{s}ci\'{n}ski, and F. Wudarski, 
Phys. Lett. A {\bf 377,} 21-22, (2013).

\bibitem{Barry}
B. M. Garraway, Phys. Rev. A {\bf 55}, 2290 (1997).

\bibitem{MazMan}
L. Mazzola, S. Maniscalco, J. Piilo, K.-A. Suominen, and B. M. Garraway, Phys. Rev. A {\bf 80}, 012104 (2009).

\bibitem{Hall2014}
M. J. W. Hall, J. D. Cresser, L.  Li, and E. Andersson, Phys. Rev. A {\bf 89}, 042120 (2014).

%\bibitem{2} H.-P. Breuer and B. Vacchini, Phys. Rev. Lett. 101 (2008) 140402; Phys. Rev. E 79, 041147 (2009).
%\bibitem{3} J. Piilo, S. Maniscalco, K. Härkönen, and K.-A. Suominen, Phys. Rev. Lett. 100, 180402 (2008).
%\bibitem{4} J. Piilo, K. Härkönen, S. Maniscalco, and K.-A. Suominen, Phys. Rev. A 79, 062112 (2009).
%\bibitem{5} W.-M. Zhang, P.-Y. Lo, H.-N. Xiong, M.W.-Y. Tu, and F. Nori, Phys. Rev. Lett. 109, 170402 (2012).
%\bibitem{6} D. Chruściński and A. Kossakowski, Phys. Rev. Lett. 104, 070406 (2010); 111, 050402 (2013).
%\bibitem{7} M. M. Wolf and J. I. Cirac, Commun. Math. Phys. 279, 147 (2008).
%\bibitem{8} A. Rivas, S. F. Huelga, and M. B. Plenio, Phys. Rev. Lett. 105, 050403 (2010).
%\bibitem{9} M. M.Wolf, J. Eisert, T. S. Cubitt, and J. I. Cirac, Phys. Rev. Lett. 101, 150402 (2008).
%\bibitem{10} S. C. Hou, X. X. Yi, S. X. Yu, and C. H. Oh, Phys. Rev. A 83, 062115 (2011).
%
%\bibitem{12} X.-M. Lu, X.Wang, and C. P. Sun, Phys. Rev. A 82, 042103 (2010).
%\bibitem{13} A. K. Rajagopal, A. R. Usha Devi, and R.W. Rendell, Phys. Rev. A 82, 042107 (2010).
%\bibitem{14} S. Luo, S. Fu, and H. Song, Phys. Rev. A 86, 044101 (2012).
%\bibitem{15} B. Bylicka, D. Chruściński, and S. Maniscalco, arXiv:1301.2585.
%\bibitem{16} S. Lorenzo, F. Plastina, and M. Paternostro, Phys. Rev. A 88, 020102 (2013).
%
%\bibitem{17} L. Mazzola, E.-M. Laine, H.-P. Breuer, S. Maniscalco, and J. Piilo, Phys. Rev. A 81, 062120 (2010).
%
%\bibitem{18} Phys. Rev. Lett. 103, 210401 (2009).
%
%\bibitem{19} Journal of Modern Optics  Volume 54, Issue 12, (2007),  arXiv:0801.4100
%
%
%\bibitem{22} A. Shabani and D. A. Lidar, Phys. Rev. A 71, 020101R (2005).


\end{thebibliography}
\end{document}